# Hydrogen trapping and interfacial decohesion at the α-$Al_2O_3$(0001)/Fe(110) interface

Youngseok Hwang, Norihito Sakaguchi, Yuji Kunisada*

Center for Advanced Research of Energy and Materials, Faculty of Engineering, Hokkaido University, Kita 13 Nishi 8 Kita-ku, Sapporo, Hokkaido 060-8628, Japan.

*Email: kunisada@eng.hokudai.ac.jp

**Abstract**

Hydrogen embrittlement and tritium leakage pose critical challenges for fusion reactor structural components, rendering α-$Al_2O_3$/Fe interfaces vital as tritium permeation barriers. Here, the thermodynamic stability, trapping energetics, and hydrogen-induced decohesion at the α-$Al_2O_3$ (0001)/Fe(110) interface were systematically investigated using density functional theory. Single-hydrogen incorporation reveals that the Fe-hollow site is the most stable trapping region, owing to local free volume and heterogeneous interfacial bonding. Multi-hydrogen analysis demonstrates that trapping behavior is concentration-dependent; increasing hydrogen concentration progressively reduces the available free volume and increases local lattice distortion. As a result, simulated cleavage processes show a monotonic decrease in cleavage energy with accumulation. At high hydrogen concentrations, cleavage energy turns negative, indicating spontaneous interfacial exfoliation. These quantitative insights clarify the atomistic degradation mechanisms of protective oxide scales, offering a theoretical framework for optimizing high-performance permeation barriers in fusion-relevant steels.

**Highlights**

-A stable α-$Al_2O_3$(0001)/Fe(110) interface model was constructed using density functional theory to evaluate tritium barrier efficacy.

-Interfacial Fe-hollow sites act as strong hydrogen traps, governed by local free volumes.

-Hydrogen trapping diminishes at high concentration due to localized lattice distortion and strain.

-Interfacial cleavage energy decreases monotonically and turns negative at high hydrogen concentrations.

-The findings provide an atomistic basis for preventing oxide-scale delamination in fusion structural materials.

## 1. Introduction

Nuclear fusion power has been widely studied as a promising solution to global energy and environmental challenges, due to its high energy density and the absence of carbon dioxide emissions during operation [1]. In the representative deuterium–tritium (D–T) fusion reaction, tritium must be bred through the nuclear reaction of lithium with neutrons generated from the fusion process within the blanket system [2]. Because tritium is radioactive, its permeation through metallic components can lead to hydrogen embrittlement. Furthermore, the leaked tritium can cause widespread environmental contamination; if it is subsequently incorporated into the human body as tritiated water (HTO), it poses a severe risk of internal radiation exposure [3].

To mitigate these issues, tritium permeation barriers (TPBs) consisting of thin dense ceramic coatings have been extensively developed. In addition to oxides, carbides and nitrides have also been considered as candidate barrier materials because of their high melting points and low hydrogen solubility [4]. However, under the corrosive and irradiation environments relevant to fusion blanket systems, carbides and nitrides may suffer from chemical instability or degradation, whereas oxide ceramics generally exhibit superior thermodynamic stability and corrosion resistance. Experimental studies have reported that $Al_2O_3$ [5,6], $Er_2O_3$ [7,8], $TiO_2$ [9], and $Cr_2O_3$ [10-12] coatings can effectively reduce hydrogen permeation by several orders of magnitude. Among these candidates, $Al_2O_3$ has attracted particular attention due to its high thermodynamic and chemical stability, corrosion resistance, and high-temperature durability. Especially, crystalline $\alpha$-$Al_2O_3$ has been predicted to exhibit a theoretical permeation-reduction factor exceeding $10^6$ [13].

FeCrAl oxide dispersion-strengthened (ODS) alloys are regarded as promising structural materials for fusion reactor blankets because pre-oxidation treatments yield a dense and adherent $\alpha$-$Al_2O_3$ scale that simultaneously provides oxidation resistance and reduced hydrogen permeability [14]. However, hydrogen permeation in such systems is not solely governed by bulk diffusion within the oxide layer; it is also influenced by interfacial phenomena. While the hydrogen transport and diffusion behavior in bulk $\alpha$-$Al_2O_3$ have been widely investigated experimentally and theoretically [15], the atomic-scale mechanisms governing hydrogen segregation, trapping energetics, and interfacial decohesion at oxide/metal interfaces remain unclear.

In various metal/oxide systems, interfaces have been identified as favorable segregation sites for hydrogen due to lattice mismatch, reduced atomic coordination, and localized electronic states. Accumulation of hydrogen at such interfaces can modify the local bonding configuration, perturb charge redistribution across the interface, and reduce the interfacial cleavage energy, thereby promoting interfacial decohesion or exfoliation. Although previous first-principles studies have examined hydrogen behavior at related oxide interfaces, including $\alpha$-$Cr_{22}O_3$/$\alpha$-$Fe_2O_3$ [16] and $\alpha$-$Al_2O_3$/$\alpha$-$Cr_2O_3$ [17], a quantitative understanding of hydrogen segregation energetics and its influence on the interfacial cohesive properties at the $\alpha$-$Al_2O_3$/FeCrAl-ODS interface is still lacking.

In the present study, we employ first-principles calculations based on density functional theory (DFT) to systematically evaluate the thermodynamic stability and segregation energetics of hydrogen at the α-$Al_2O_3$(0001)/α-Fe(110) interface. As a fundamental step toward understanding hydrogen behavior at the α-$Al_2O_3$/FeCrAl-ODS interface, we focus on the interface formed between the most stable surfaces of each constituent, namely α-$Al_2O_3$(0001) and α-Fe(110). This simplified model system can focus on the essential interfacial bonding characteristics while minimizing additional complexities arising from alloying elements and microstructural heterogeneity. The present model should, therefore, be regarded as a first-step approximation toward the more complex α-$Al_2O_3$/FeCrAl interface. A stable interface model with minimal lattice mismatch is first constructed and fully relaxed. Hydrogen impurity is subsequently introduced into symmetry-inequivalent interfacial sites to determine the preferred interstitial sites and the corresponding formation energies. In addition, we investigate the effect of interfacial hydrogen impurities on the mechanical stability of the $Al_2O_3$/Fe system by calculating the work of exfoliation with and without hydrogen incorporation, thereby elucidating the extent to which hydrogen impurities weaken interfacial bonding and promote interfacial exfoliation of the protective $Al_2O_3$ scale from the Fe substrate.

## 2. Calculation methods

First-principles calculations based on DFT were performed to determine the energetically most favorable configuration and the electronic structure of the α-$Al_2O_3$/Fe interface, as well as the thermodynamic stability of hydrogen impurities incorporated at the interface.

All calculations were carried out using the Vienna Ab initio Simulation Package (VASP) [18–21], which employs plane waves as basis functions within the projector augmented-wave (PAW) framework. The generalized gradient approximation (GGA) proposed by Perdew, Burke, and Ernzerhof (PBE) was adopted for the exchange–correlation functional [22]. The Monkhorst–Pack scheme [23] was used for Brillouin-zone sampling, with Gaussian smearing of 0.1 eV.

The cutoff energy for the plane-wave basis was set to 600 eV. A Γ-centered 1 × 1 × 1 k-point mesh was employed for the interfacial supercell, which ensured total-energy convergence within 0.01 eV/atom. Atomic positions were fully relaxed until the Hellmann–Feynman forces acting on each atom became less than 0.02 eV/Å.

Charge transfer between atoms was analyzed using the Bader charge partitioning scheme [24–26]. The crystal orbital Hamilton population (COHP) analysis was performed with the Local Orbital Basis Suite Towards Electronic-Structure Reconstruction (LOBSTER) code developed by Maintz and Dronskowski to evaluate the chemical bonding characteristics between orbitals [27,28]. Atomic structures and charge-density distributions were visualized using the Visualization for Electronic and STructural Analysis (VESTA) software [29].

We constructed the interfacial supercell by connecting the Fe and α-$Al_2O_3$ slabs, which contains two interfaces. In this study, we considered only the Al-terminated α-$Al_2O_3$(0001) slab in constructing the interface model, based on the previous DFT study where the Al-terminated (0001) surface shows the lowest surface energy.[30] In addition, we can construct the stoichiometric α-$Al_2O_3$ slab. As for bcc-Fe surface, we adopted the well-known most stable Fe(110) slab.

To construct the interfacial supercell, we considered the rectangular surface unit cell of α-$Al_2O_3$(0001)-$(1 \times \sqrt{3})$ used as a reference for lattice matching with Fe(110), as shown in Fig. 1. The calculated equilibrium surface lattice constants of α-$Al_2O_3$ ($[\bar{1}\ 2\ \bar{1}\ 0]$:4.805 Å, $[\bar{1}\ 0\ 1\ 0]$:8.323 Å) and bcc-Fe ($[0\ 0\ 1]$:2.830 Å, $[1\ \bar{1}\ 0]$:4.002 Å) were consistent with previous experimental reports [31,32]. To minimize the lattice mismatch between the two crystals, we adopted 3×1 α-$Al_2O_3$(0001) slab and a 5×2 Fe(110) slab. The Fe lattice was slightly stretched to fit the α-$Al_2O_3$ lattice, introducing in-plane tensile strains of about 1.9% and 4.0% along the $[0\ 0\ 1]$ and $[1\ \bar{1}\ 0]$ directions, respectively. After the α-$Al_2O_3$ and Fe slabs were brought into contact, the interfacial structure was fully relaxed by optimizing all atomic coordinates and the supercell dimension normal to the interface plane. The slab thicknesses were determined from systematic convergence tests and selected such that the interface energy was converged to within 1%. Finally, we adopted the interfacial supercell shown in Fig. 2, possesses an equilibrium cell length of 21.923 Å along the interface-normal direction. The corresponding interface energy, obtained from the excess energy of the interfacial system with respect to the constituent bulk phases, was found to be 3.61 J/$m^2$. This obtained supercell size was fixed in all subsequent calculations.

After hydrogen incorporation into the interfacial model, the atomic structure was fully relaxed. The hydrogen formation energy at the interface, $E_{\mathrm{f}}$, was defined as

$$E_{\mathrm{f}} = E_{\mathrm{interface}}^{\mathrm{H}} - \left(E_{\mathrm{interface}} + \frac{1}{2}E_{\mathrm{H_2}}\right), \quad (1)$$

where $E_{\mathrm{interface}}^{\mathrm{H}}$ and $E_{\mathrm{interface}}$ are the total energies of the interfacial supercell with and without hydrogen, respectively. $E_{\mathrm{H_2}}$ denotes the total energy of an isolated $H_2$ molecule calculated in a 20 Å cubic cell to eliminate intermolecular interactions between periodic cells.

To evaluate the effect of hydrogen impurities on the mechanical stability of the interface, ideal cleavage processes with and without hydrogen were simulated by separating the α-$Al_2O_3$/Fe interface along the direction normal to the interface plane. This procedure represents an ideal brittle exfoliation without considering external loading or crack propagation effects. The interfacial cohesive strength was quantified from the total-energy difference between the bonded interface model (before cleavage) and the separated slab models of α-$Al_2O_3$ and Fe (after cleavage). The cleavage energy of the clean interface, $\gamma_{\mathrm{c}}^{\mathrm{clean}}$, was defined as

$$\gamma_{\mathrm{c}}^{\mathrm{clean}} = \frac{1}{2A}\left(E_{\mathrm{slab}}^{\mathrm{Fe}} + E_{\mathrm{slab}}^{\alpha\text{-}\mathrm{Al_2O_3}} - E_{\mathrm{interface}}\right), \quad (2)$$

where $A$ denotes the interfacial area. $E_{\mathrm{slab}}^{\mathrm{Fe}}$ and $E_{\mathrm{slab}}^{\alpha\text{-}\mathrm{Al_2O_3}}$ are the total energies of the isolated Fe and α-$Al_2O_3$ slabs, respectively. For the hydrogen-containing system, the interfacial model consists of two interfaces, therefore, with and without hydrogen impurities. The cleavage energy of the interface with hydrogen impurities, $\gamma_c^{\mathrm{H}}$, was obtained as follows:

$$\gamma_{\mathrm{c}}^{\mathrm{H}} = \frac{1}{A}\left(E_{\mathrm{slab}}^{\mathrm{Fe+H}} + E_{\mathrm{slab}}^{\alpha\text{-}\mathrm{Al_2O_3}} - E_{\mathrm{interface}}^{\mathrm{H}}\right) - \gamma_{\mathrm{c}}^{\mathrm{clean}}, \quad (3)$$

where $E_{\mathrm{slab}}^{\mathrm{Fe+H}}$ denotes the total energy of the Fe slab with hydrogen retained on the Fe surface after cleavage. Hydrogen remaining on the α-$Al_2O_3$ surface after exfoliation was found to be energetically unfavorable; therefore, only configurations in which hydrogen resides on the Fe surface were considered in the evaluation of cleavage energetics.

## 3. Results and discussion

In this study, we considered four hydrogen incorporation sites. The subsurface layers of Fe and α-$Al_2O_3$ are denoted as Fe-sub and α-$Al_2O_3$-sub, respectively. Sites located directly beneath oxygen atoms are referred to as O-top sites, whereas hollow sites surrounded by interfacial Fe atoms are denoted as Fe-hollow sites in the interface region. Owing to the incoherent nature of the α-$Al_2O_3$(0001)/Fe(110) interface, hydrogen formation energies were evaluated for multiple symmetry-inequivalent interstitial sites at the α-$Al_2O_3$/Fe interface.

Figure 3 summarizes the hydrogen formation energies obtained after structural relaxation. Independent of the initial position, hydrogen atoms introduced into the interfacial region preferentially relaxed toward either Fe-hollow or O-top sites. Among these configurations, the Fe-hollow site exhibited the lowest formation energy, indicating that it serves as the most favorable hydrogen-trapping site at the α-$Al_2O_3$/Fe interface. The minimum formation energy was calculated to be −0.44 eV, showing that hydrogen is stabilized at the interface. In contrast, hydrogen located on the α-$Al_2O_3$ side was considerably less stable. Even among Fe-hollow sites, the formation energy varied by approximately 0.6 eV, suggesting that the local atomic environment affects hydrogen stability.

To assess the trapping capability of the interface, the present results were compared with previously reported hydrogen formation energies in bulk α-$Al_2O_3$ and bcc-Fe, which are approximately 3.51 eV and 0.30 eV, respectively.[33,34] As shown in Fig. 4, the hydrogen formation energies at most interfacial sites are substantially lower than those in the corresponding bulk phases. This trend indicates that the α-$Al_2O_3$/Fe interface acts as a trapping region for hydrogen impurities. As a result, hydrogen atoms that diffuse through the oxide layer are expected to accumulate at the interface.

To clarify the origin of the distribution of hydrogen-trapping strength at Fe-hollow sites, the local atomic environments around corresponding sites were analyzed. Figure 5 shows four characteristic interfacial configurations of Fe-hollow sites, denoted as sites A–D. Each site corresponds to the center of a triangular arrangement formed by three interfacial Fe atoms. Sites A and B are Fe-hollow sites that do not contain neighboring Al atoms, whereas site D is located directly beneath an interfacial Al atom. Site C corresponds to an intermediate bridge-like configuration, in which an interfacial Al atom is located above an Fe–Fe bridge site.

Hydrogen stability strongly depends on the local geometry of these sites. Site D exhibits the smallest available free volume owing to the short Al–Fe separation, making hydrogen unstable. Although hydrogen can occupy site C, its stability is limited by the restricted free volume. In contrast, sites A and B provide larger free volumes and therefore exhibit lower formation energies. The difference between sites A and B originates from the local bonding environment of the surrounding Fe atoms. In site B, one of the three Fe atoms forming the Fe-hollow site does not participate in interfacial bonding with Al atoms. As a result, this Fe atom is displaced toward the oxide side of the interface, resulting in a reduced free volume compared with site A. The smaller free volume increases the local lattice strain associated with hydrogen incorporation, leading to a slightly higher formation energy at site B than at site A.

To quantitatively evaluate the effect of local geometry, Voronoi volumes were calculated as free volumes for each hydrogen-trapping site. As shown in Fig. 6, the Voronoi volume generally follows the order A > B > C, consistent with the observed trend in hydrogen formation energy. Larger free volumes provide a more favorable environment for hydrogen impurities, resulting in enhanced thermodynamic stability. Minor deviations from this correlation indicate that factors other than free volume, such as the local chemical environment and H–Al interactions, also contribute to hydrogen stability.

To further elucidate the electronic origin of the distribution of hydrogen-trapping strength, charge-density redistribution at the α-$Al_2O_3$/Fe interface was analyzed. Figure 5 also presents the differential charge density of the α-$Al_2O_3$/Fe interface. Charge accumulation is observed between neighboring Al and Fe atoms with separations shorter than approximately 3 Å, indicating the formation of Al–Fe bonds. This interpretation is also supported by the COHP analysis, which indicates that the bonding energy ranges from −0.8 to −2.8 eV. In contrast, Fe atoms located away from Al atoms participate weakly in interfacial bonding, resulting in spatially heterogeneous bonding environments due to interfacial incoherency. Such heterogeneity gives rise to interstitial sites with different free volumes and bonding strengths, thereby explaining the wide distribution of hydrogen formation energies observed at the interface.

The preceding results demonstrate that the α-$Al_2O_3$/Fe interface acts as a trapping region for isolated hydrogen impurities. Under practical operating conditions, however, multiple hydrogen impurities may accumulate at the interface. To investigate the concentration dependence of hydrogen stability, various multi-hydrogen configurations were constructed using the stable sites A and B identified in the single-hydrogen systems.

Table 1 summarizes the average hydrogen formation energy per atom $\overline{E}_{\mathrm{f}}$, the value estimated from isolated hydrogen calculations $\overline{E_{\mathrm{f}}^{est}}$, and the corresponding H–H separation distance $d_{\mathrm{H-H}}$. The comparison between $\overline{E}_{\mathrm{f}}$ and $\overline{E_{\mathrm{f}}^{est}}$ provides insight into the interaction between neighboring hydrogen atoms at the interface. For well-separated hydrogen atoms with H–H separation longer than 2 Å, the calculated average formation energy is comparable to or slightly lower than the estimated value. These results indicate that hydrogen impurities behave independently when sufficient separation is maintained and that the interfacial region can accommodate multiple hydrogen impurities without a significant energetic penalty.

In contrast, configurations containing neighboring hydrogen atoms, i.e., H–H separations shorter than 2 Å, exhibit higher formation energies than the values estimated from isolated-hydrogen calculations. This destabilization reflects the overlap of local lattice distortions and the repulsive interaction associated with closely spaced hydrogen impurities. As hydrogen concentration increases further, all A, all B, and all AB configurations show that the energetic advantage of interfacial trapping diminishes. This trend suggests that the finite interfacial free volume is consumed by hydrogen incorporation, leading to increased local strain and reduced trapping efficiency. In brief, the α-$Al_2O_3$/Fe interface acts as an efficient hydrogen trap under dilute conditions. However, hydrogen stability depends on its concentration, and the thermodynamic driving force for hydrogen trapping decreases as hydrogen accumulates at the interface.

To evaluate the influence of hydrogen accumulation on interfacial mechanical stability, cleavage energies were calculated for hydrogen-containing α-$Al_2O_3$/Fe interfaces. The calculated results are summarized in Table 2, and the concentration dependence, mainly based on the results from site A, is illustrated in Fig. 7. In this study, hydrogen concentration is evaluated by the hydrogen coverage (ML) at the interfacial Fe-hollow sites. The cleavage energy decreases monotonically with increasing hydrogen coverage. While isolated hydrogen atoms have only a minor influence on interfacial bond strength, progressive hydrogen accumulation leads to a significant reduction in the energy required to separate the interface. At high hydrogen concentration, the cleavage energy becomes negative, indicating that the separated-slab configuration, i.e., the exfoliated state, is more stable than the bonded interface.

A notable deviation from the nearly linear concentration dependence is observed for all B configurations. As discussed, B-type sites possess smaller Voronoi volumes and exhibit local distortion due to the displacement of neighboring Fe atoms toward the oxide side. Simultaneous occupation of all B sites, therefore, results in enhanced overlap of hydrogen-induced lattice relaxation, causing an abrupt decrease in cleavage energy compared with neighboring configurations.

These results demonstrate that the α-$Al_2O_3$/Fe interface plays an important role in hydrogen transport. Under dilute conditions, the interface serves as an effective hydrogen-trapping region. However, continued hydrogen accumulation progressively weakens interfacial bonding and promotes decohesion. At high hydrogen concentrations, the interface becomes unstable against exfoliation, suggesting that hydrogen-assisted exfoliation of the protective α-$Al_2O_3$ scale may occur under severe hydrogen exposure conditions or during long-term operation. It should be noted that the highest hydrogen coverages considered in the present study represent idealized limiting cases designed to elucidate the influence of hydrogen accumulation on interfacial cohesion. Whether such hydrogen concentrations can be achieved under actual fusion reactor operating conditions depends on hydrogen flux, diffusion kinetics, trapping characteristics, irradiation-induced defects, and service lifetime. Therefore, the present results should be interpreted as revealing the thermodynamic tendency of hydrogen-assisted interfacial decohesion rather than predicting a specific hydrogen concentration for oxide-scale exfoliation.

## 4. Conclusion

In this study, the structural stability, hydrogen-trapping behavior, and hydrogen-induced decohesion of the α-$Al_2O_3$(0001)/Fe(110) interface were systematically investigated using first-principles calculations based on DFT. The optimized Al-terminated α-$Al_2O_3$(0001)/Fe(110) interface exhibited a low interface energy of 3.61 J/$m^2$, indicating a stable interfacial structure.

Hydrogen incorporation calculations revealed that the Fe-hollow site is the most favorable trapping site at the interface, with a formation energy of −0.44 eV. This value is substantially lower than those reported for bulk Fe and α-$Al_2O_3$, demonstrating that the interface acts as an effective hydrogen-trapping region. Analysis of the local atomic structure, Voronoi volume, and charge redistribution showed that hydrogen stability is governed by both the available free volume and the heterogeneous interfacial bonding environment.

The stability of interfacial hydrogen was found to be dependent on hydrogen concentration. While the interface can accommodate multiple hydrogen atoms under dilute conditions, increasing hydrogen coverage progressively reduces the available free volume and enhances local lattice distortion, thereby diminishing the energetic advantage of hydrogen trapping.

Calculated cleavage energy demonstrated that isolated hydrogen atoms have only a minor influence on interfacial cohesion. In contrast, hydrogen accumulation significantly weakens interfacial bonding, resulting in a monotonic decrease in cleavage energy with increasing hydrogen coverage. At high hydrogen concentrations, the cleavage energy becomes negative, indicating that hydrogen accumulation can render the interface unstable against exfoliation and may promote spontaneous exfoliation of the protective oxide scale. In brief, the α-$Al_2O_3$/Fe interface acts as a strong hydrogen-trapping site under dilute conditions, whereas hydrogen accumulation significantly reduces interfacial cohesion and may induce interfacial exfoliation.

The present study focused on the stable Al-terminated α-$Al_2O_3$(0001)/Fe(110) interface as a model system. However, the surface termination and atomic structure of corundum oxides are known to be sensitive to factors such as oxygen chemical potential, temperature, and processing conditions.[35] Future studies should therefore investigate hydrogen behavior at interfaces involving different α-$Al_2O_3$ terminations and crystallographic orientations, as well as more realistic oxide/metal interface structures with an additive element in Fe-based materials. Such investigations will provide a more comprehensive understanding of hydrogen trapping and hydrogen-assisted degradation mechanisms in oxide-coated fusion structural materials.

**Declaration of the use of generative AI and AI-assisted technologies**

During the preparation of this work, the authors used Google Gemini 3.5 solely for English language editing, including grammar correction and spelling checks. The authors reviewed and edited all suggested changes and take full responsibility for the content of the publication.

**Declaration of competing interest**

The authors declare that they have no competing financial interests or personal relationships that could have appeared to influence the work reported in this paper.

**Acknowledgments**

This work has been done using the facilities of the Supercomputer Center, the Institute for Solid State Physics, the University of Tokyo.

Table 1 Average hydrogen formation energy per atom ($\overline{E}_{\mathrm{f}}$), estimated formation energy derived from isolated single-hydrogen values ($\overline{E_{\mathrm{f}}^{\mathrm{est}}}$), and H–H separation distance ($d_{\mathrm{H-H}}$) for multiple-hydrogen configurations at the α-$Al_2O_3$/Fe interface.

| Configuration | $\overline{E}_{\mathrm{f}}$ (eV) | $\overline{E_{\mathrm{f}}^{\mathrm{est}}}$ (eV) | $d_{\mathrm{H-H}}$ (Å) |
|---|---|---|---|
| A–A | −0.438 | −0.421 | 5.054 |
| A–A | −0.227 | −0.372 | 1.700 |
| A–A | −0.292 | −0.347 | 1.835 |
| A–A | −0.354 | −0.411 | 1.838 |
| A–A | −0.321 | −0.391 | 1.860 |
| A–B | −0.281 | −0.203 | 3.724 |
| A–B | −0.348 | −0.338 | 4.201 |
| A–B | −0.192 | −0.337 | 1.666 |
| A–B | −0.278 | −0.277 | 2.219 |
| all A | −0.276 | −0.364 | – |
| all B | −0.186 | −0.216 | – |
| all AB | −0.107 | −0.273 | – |

Table 2 Hydrogen-concentration dependence of cleavage energy ($\gamma_c$). Hydrogen concentration is evaluated by the hydrogen coverage (ML) at the interfacial Fe-hollow sites, and the change in cleavage energy is referenced to the clean-interface value. Values in parentheses in the first column represent the H–H separation distances.

| Configuration | Number of H atoms | Coverage (ML) | $\gamma_c$ (J/m$^2$) | Cleavage energy change (%) |
|---|---|---|---|---|
| Clean | 0 | 0.000 | 0.718 | |
| Fe-top | 1 | 0.025 | 0.697 | −2.9 |
| Fe-hollow | 1 | 0.025 | 0.686 | −4.4 |
| A-A (5.05 Å) | 2 | 0.050 | 0.668 | −7.0 |
| A-A (1.70 Å) | 2 | 0.050 | 0.654 | −9.0 |
| all A | 11 | 0.275 | 0.412 | −42.6 |
| all B | 18 | 0.450 | −0.011 | −101.6 |
| all AB | 29 | 0.725 | −0.102 | −114.2 |

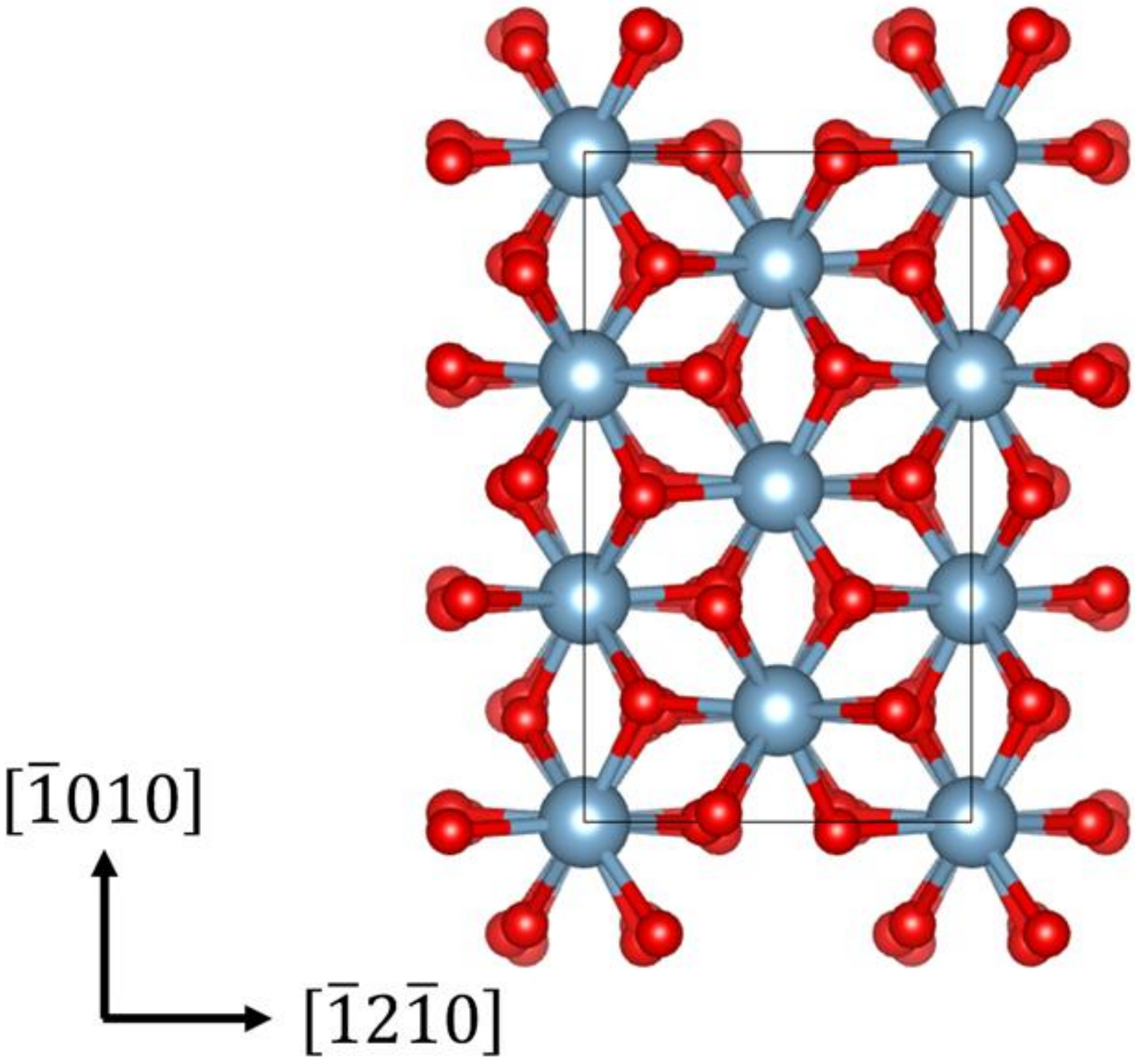

Fig. 1 Rectangular surface unit cell of α-$Al_2O_3$(0001)-$(1 \times \sqrt{3})$ surface. The blue and red spheres represent Al and O atoms, respectively.

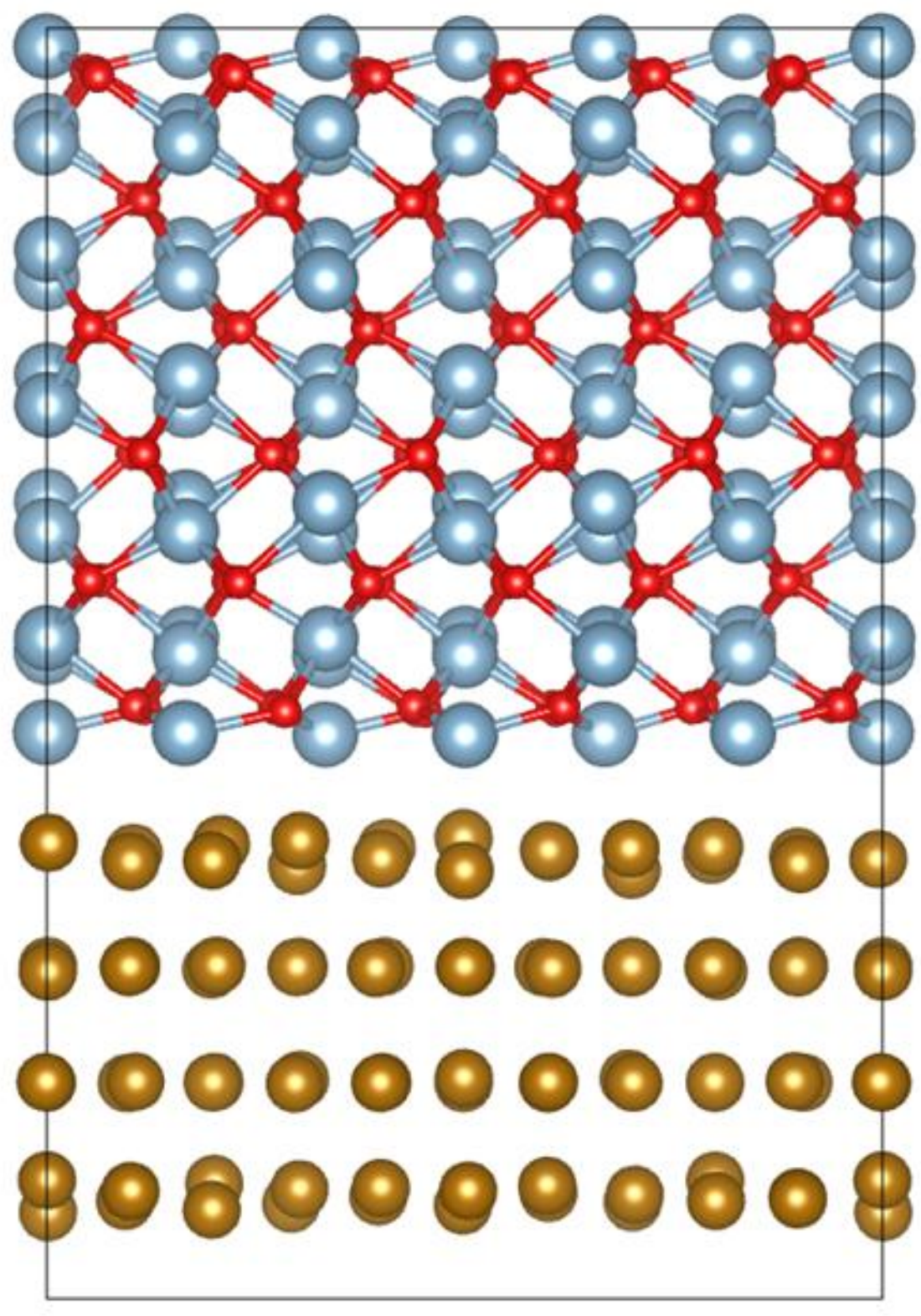

Fig. 2 Constructed α-$Al_2O_3$/Fe interface model consisting of the Al-terminated α-$Al_2O_3$(0001) surface and the Fe(110) surface. The blue, red, and brown spheres represent Al, O, and Fe atoms, respectively.

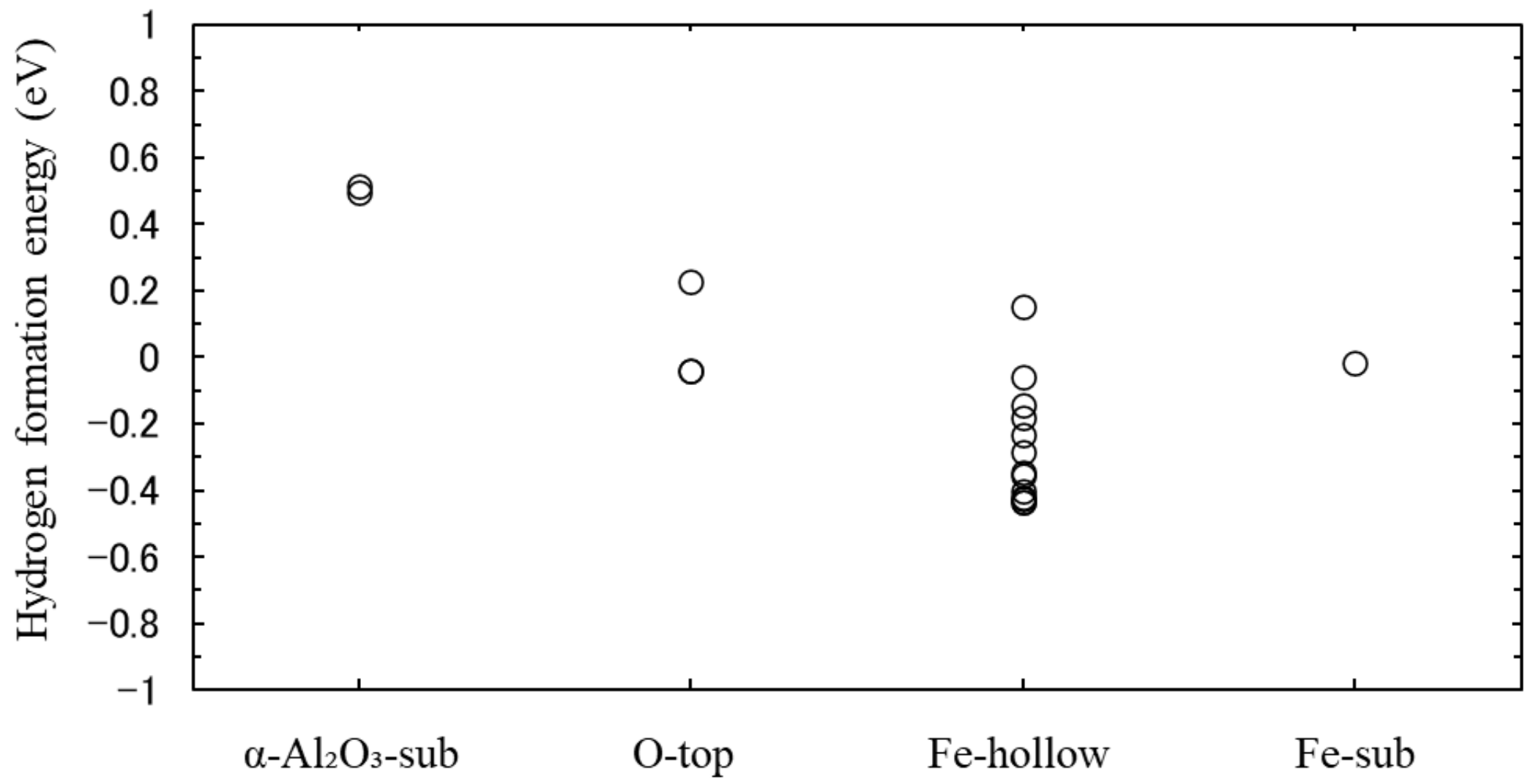

Fig. 3 Hydrogen formation energies after structural relaxation at the α-$Al_2O_3$/Fe interface.

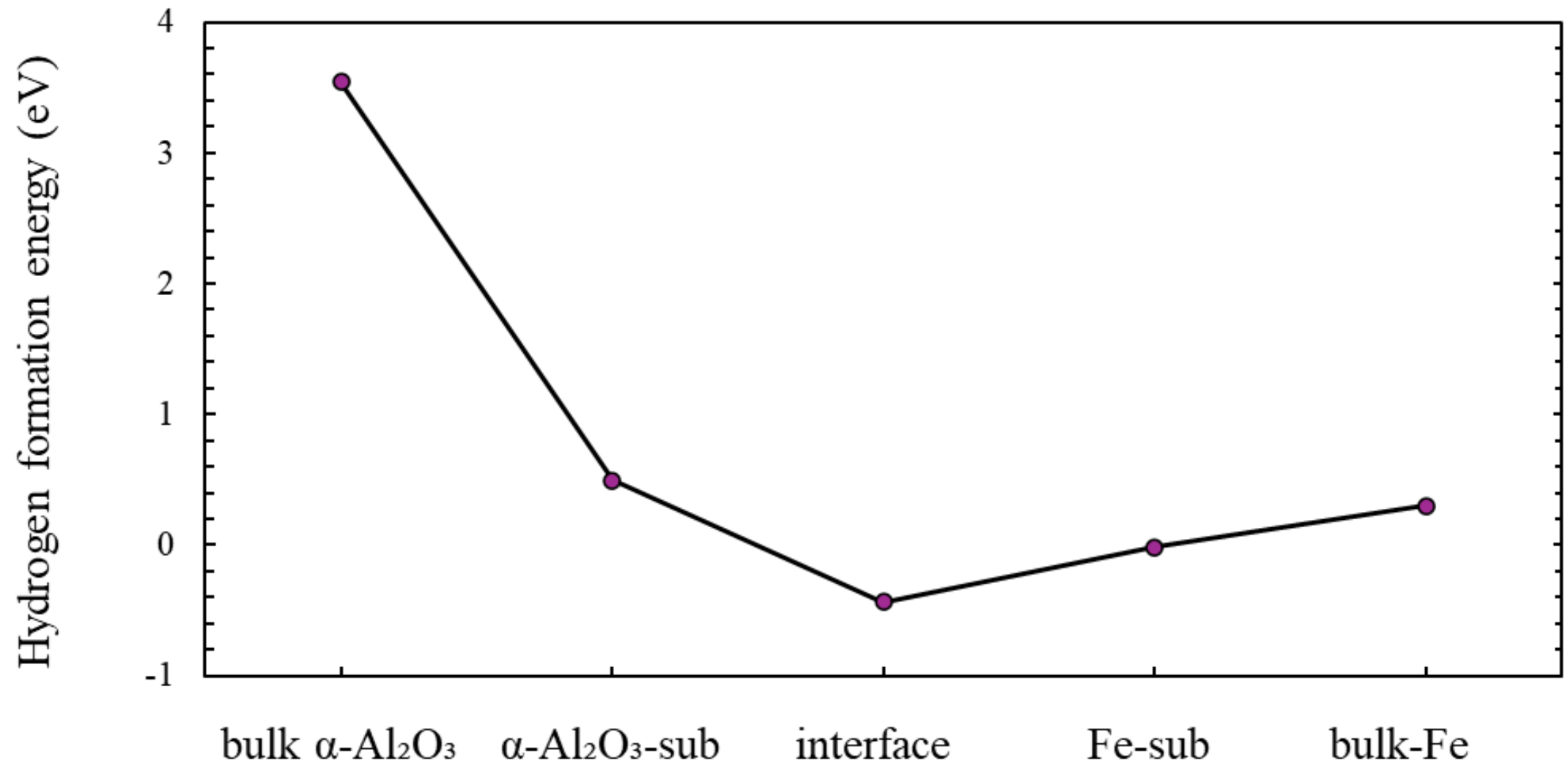

Fig. 4 Comparison of hydrogen formation energies among bulk Fe, bulk α-$Al_2O_3$ [33, 34], and the α-$Al_2O_3$/Fe interface obtained in this study.

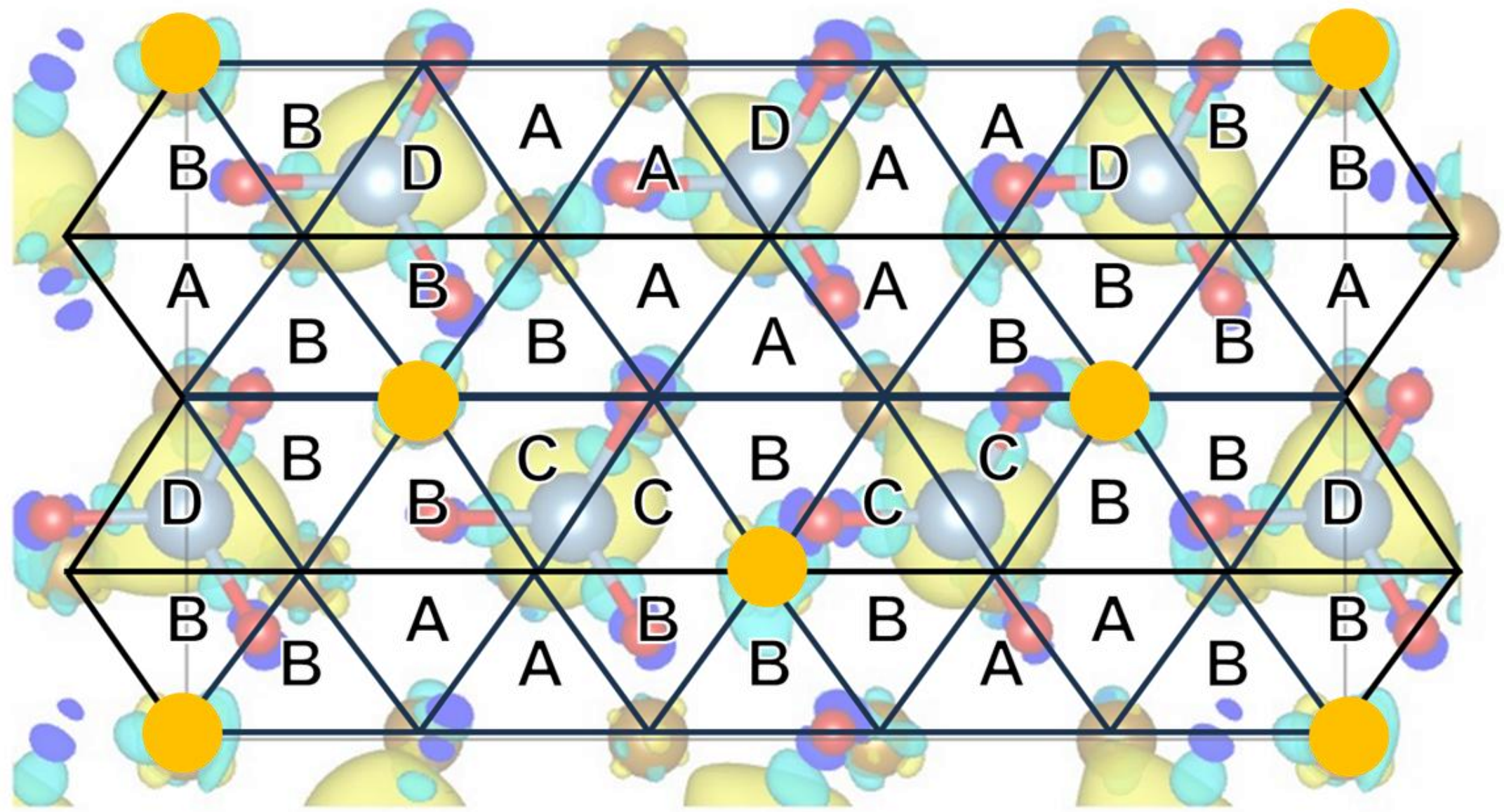

Fig. 5 Schematic illustration of the α-$Al_2O_3$/Fe interface showing the representative hydrogen-trapping sites (A–D). Each site corresponds to the center of a triangular lattice formed by three interfacial Fe atoms. The yellow and blue regions indicate electron accumulation and depletion (±0.05 e/$Å^3$), respectively. The orange circles indicate interfacial Fe atoms that do not form bonds with neighboring Al atoms.

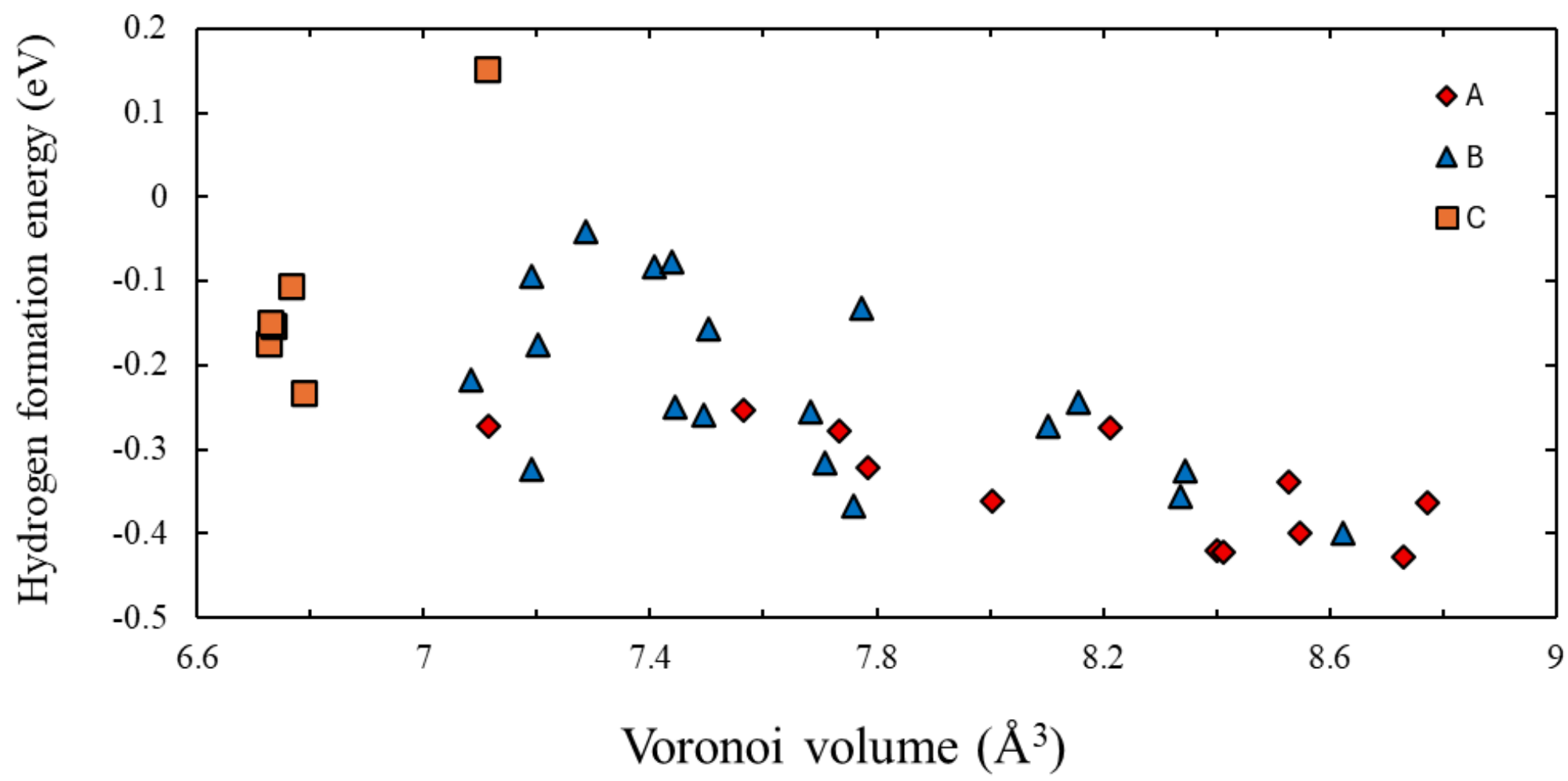

Fig. 6 Voronoi volumes of hydrogen-trapping sites (A, B, and C) at the α-$Al_2O_3$/Fe interface.

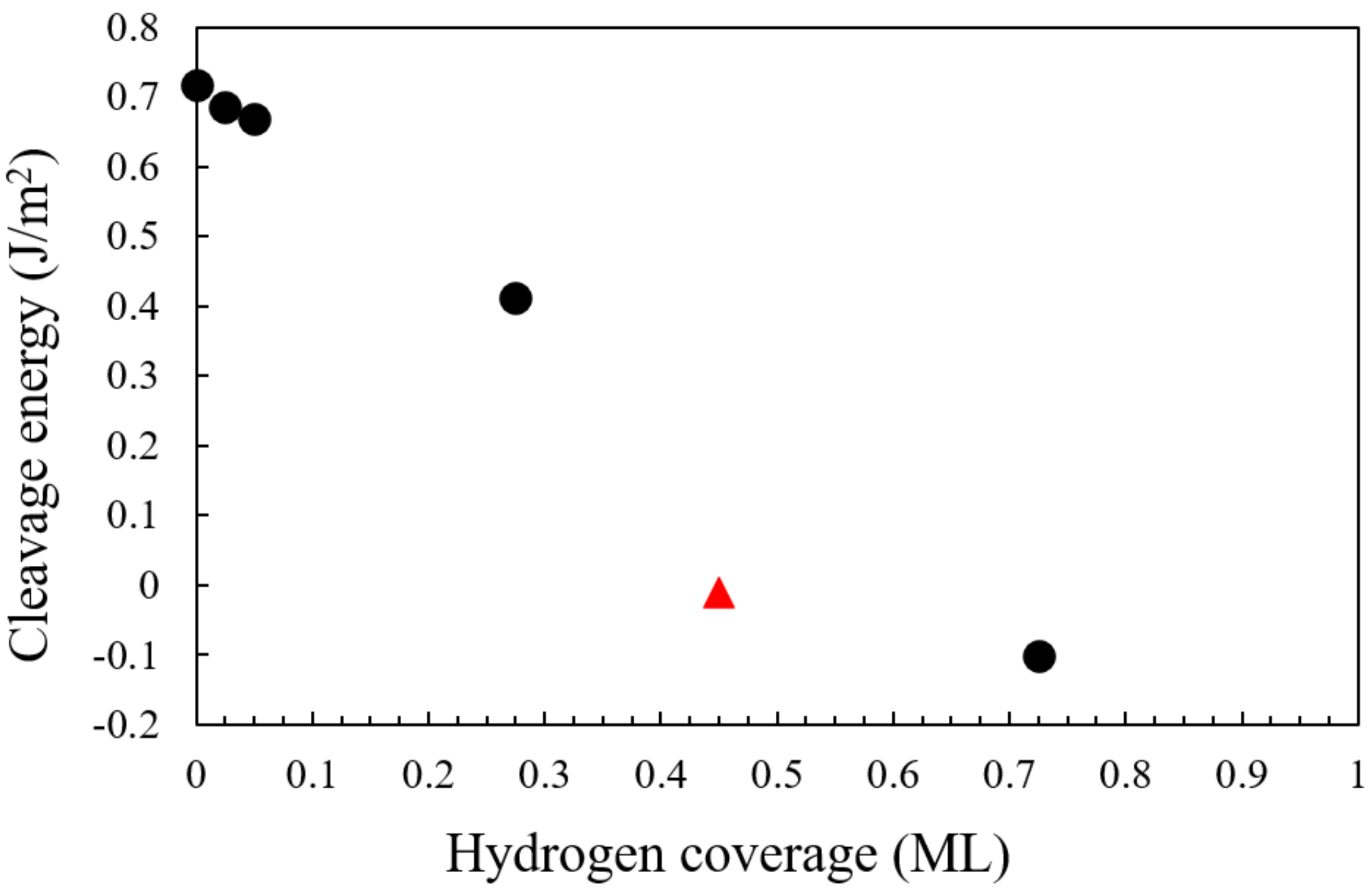

Fig. 7 Cleavage energy as a function of hydrogen coverage at the α-$Al_2O_3$/Fe interface. The black circles represent the results for site A and all AB, and the red triangle represents the result for all B.